\documentclass[aps, prd, tightenlines, letterpaper, amsmath, amssymb, preprintnumbers, floatfix, longbibliography, nofootinbib, twocolumn]{revtex4-2}

\usepackage{latexsym,amsfonts,bm}
\usepackage{xspace}
\usepackage{cancel}
\usepackage{graphicx}    % Graphics handling of JPEG, PNG, PDF
\usepackage{microtype}   % Finetuned typesetting control
\usepackage{amsmath}     % AMS Math package; equation environments, math addons
\usepackage{amsthm}    % AMS theorem-like environments
\usepackage{amssymb}   % AMS math symbols and styles like \mathbb/cal/frak/... (loads amsfonts)
\usepackage{siunitx}     % Unified number/unit handling
\usepackage{booktabs}    % Better table control
\usepackage{hyperref}
\hypersetup{colorlinks=true,citecolor=blue,linkcolor=blue}
\usepackage[capitalize]{cleveref}
\usepackage{tensor}
\usepackage[normalem]{ulem}

\usepackage{adjustbox}    % Circuit size control
\usepackage{subcaption}
\usepackage[dvipsnames]{xcolor}
\usepackage{mathtools} % Fine-tune math typesetting
\usepackage{bm}
\usepackage{dsfont}
\usepackage[shortlabels]{enumitem}  % Customize list envronments
\usepackage{physics}   % Useful bracket/vector/etc. macros
\usepackage{orcidlink}

\providecommand{\ie}{\emph{i.e.} }

\providecommand{\CN}{\mathcal{N}}

\newcommand{\SubFigRef}[2]{\ref{#1}{\color{blue}{#2}}}

\newcommand{\mbs}{s}
\newcommand{\mbk}{k}
\newcommand{\ps}{\text{p}(s)}
\newcommand{\psk}[1]{\text{p}(s; #1)}

\usepackage{accents}
\usepackage{booktabs}
\usepackage{tabularx}

\begin{document}

\begin{figure}
  \vskip -1.cm
  \leftline{\includegraphics[width=0.15\textwidth]{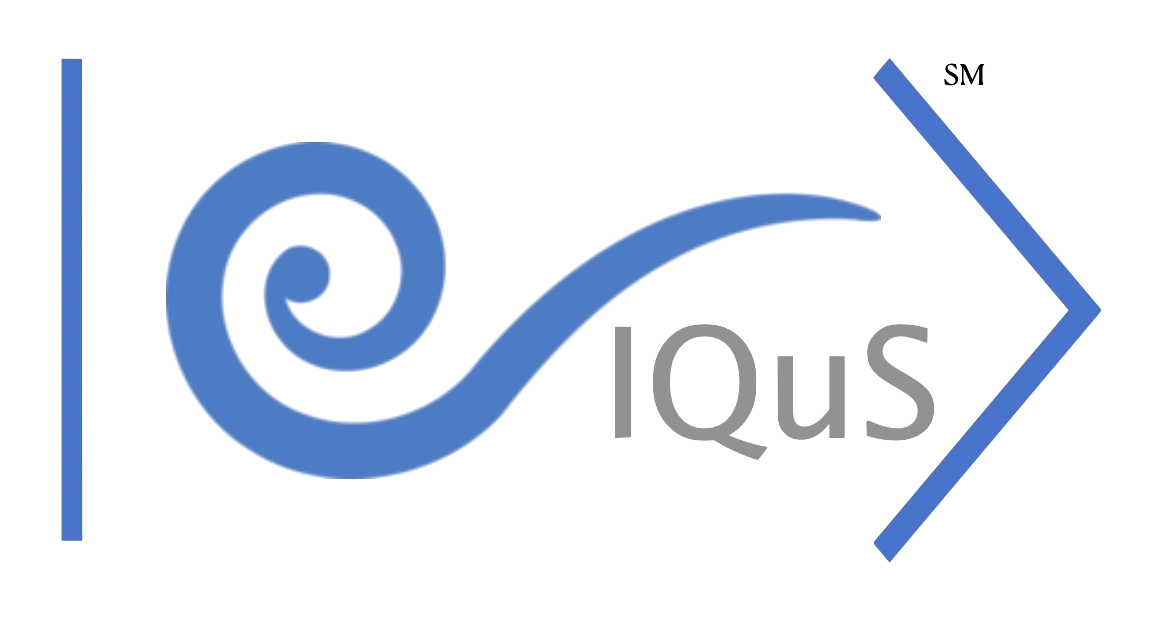}}
\end{figure}

\title{Measuring Non-Stabilizerness in an SU(2) Lattice Gauge Theory}

\author{Henry Froland\,\orcidlink{0009-0008-4356-0602}}
\email{frolandh@uw.edu}
\affiliation{InQubator for Quantum Simulation (IQuS), Department of Physics, University of Washington, Seattle, WA 98195}

\author{Dorota M. Grabowska\,\orcidlink{0000-0002-0760-4734}}
\email{grabow@uw.edu}
\affiliation{InQubator for Quantum Simulation (IQuS), Department of Physics, University of Washington, Seattle, WA 98195}

\preprint{IQuS@UW-21-129}
\date{\today}

\begin{abstract}
One of the goals of quantum simulation is to provide novel insights into quantum systems, such as the gauge theories that are relevant for high-energy and nuclear physics. Recent years have seen rapid improvements in both the hardware and software necessary for these simulations. A central consideration in the design of such simulations is the quantum complexity of a given quantum state. This work takes a step towards studying a specific kind of complexity, namely the non-stabilizerness, in a simple yet non-trivial system: SU(2) lattice gauge theory  of two plaquettes. The non-stabilizerness of low-energy eigenstates is studied and the implications for quantum simulations are discussed. The real-time evolution of this system is simulated on  \textbf{ibm\_marrakesh} and the non-stabilizerness is measured using a random measurement protocol. New techniques enhancing the efficiency of this protocol are developed, including both a new way to calculate the estimator for non-stabilizerness and a flexible error mitigation technique called Bit String Decoherence Renormalization. This mitigation method is central to accurately resolving the experimental time dependence of non-stabilizerness, and is anticipated to have broad applicability in digital quantum simulations.
\end{abstract}

\maketitle

\newpage
\section{Introduction}
Applications of quantum computers to simulating particle and nuclear physics have attracted considerable interest in recent years, enabling \emph{ab initio} studies of key phenomena such as particle collisions, confinement, and thermalization~\cite{Benioff1980,Feynman1982Simulating,feynman1985,lloyd1996universal,Bauer:2019qxa,Bauer:2023qgm,DiMeglio:2023nsa,banuls2020simulating,Bauer:2025nzf,Chai:2023qpq,Angelides:2023noe,Klco:2018kyo,Farrell:2023fgd,Farrell:2024fit,Farrell:2025nkx,Chernyshev:2025lil,mueller2025quantum,de2024observation,Zemlevskiy:2024vxt}. Gauge theories play a central role in the Standard Model and so their lattice regularizations, known as Lattice Gauge Theories (LGT), have been a primary area of focus~\cite{Halimeh:2025vvp,Byrnes:2005qx,Zohar:2015hwa,Martinez:2016yna,Zache:2018jbt,Mil:2019pbt,Klco:2019evd,Yang:2020yer,Zhou:2021kdl,Ciavarella:2021nmj,Davoudi:2022xmb,Banerjee:2012pg,Ott:2020ycj,cochran2024visualizing,Atas:2021ext,Atas:2022dqm,Ciavarella:2021lel,Buser:2020cvn,Alexandru:2019nsa,Riechert:2021ink,Ciavarella:2024fzw,Farrell:2022vyh,Farrell:2022wyt,Halimeh:2024bth,Bergner:2024qjl,Balaji:2025afl,Schuhmacher:2025ehh,Osborne:2022jxq}. These are paradigmatic models of strongly-coupled quantum many-body systems and describe a wide array of physical systems, making them ideal targets for future quantum simulators.

Much of the difficulty in understanding the properties of strongly-coupled systems stems from their underlying quantum complexity properties. Initially, quantum complexity was understood through the lens of entanglement, a mechanism that allows for correlations in systems beyond what is possible classically~\cite{vidal2003efficient,amico2008entanglement,eisert2008area,kaufman2016quantum, rakovszky2019signatures}. However, it has been realized that entanglement is not the only property responsible for the observed complexity in quantum systems, with non-stabilizerness~\cite{gottesman1998heisenberg,howard2017application,leone2022stabilizer,oliviero2022measuring,gu2025magic,tirrito2024quantifying}, or ``magic", being another essential resource. Non-stabilizerness quantifies the amount of non-Clifford resources needed to prepare a particular state. While first viewed as a quantifier of computational complexity, the time evolution of non-stabilizerness has been linked to physical phenomena like quantum chaos and thermalization~\cite{leone2021quantum,tirrito2024anticoncentration,turkeshi2025magic}. More recently, its relevance to high-energy physics has been explored in contexts like particle interactions~\cite{robin2026quantum,robin2025quantum}, the AdS-CFT correspondence~\cite{cao2025gravitational} and string-breaking ~\cite{grieninger2026quantum,grieninger2026nonlocal}. However, systematic studies of the role of non-stabilizerness in the dynamical properties of non-abelian LGTs remain relatively unexplored. 

One additional subtlety when studying non-stabilizerness in the context of LGTs is that the Hilbert space of the gauge field is formally infinite dimensional. Non-stabilizerness is naturally defined on qudit degrees of freedom, which have finite local dimension. Therefore to both simulate such theories and make contact with standard measures of non-stabilizerness, it is necessary to ``digitize" the Hilbert space. This renders the gauge field finite so that it may be mapped onto a register of qudits. There is no unique way to perform this mapping, and a variety of different bases have been developed~\cite{Mathur:2004kr,Raychowdhury:2018osk,Davoudi:2020yln,Mathur:2015wba, DAndrea:2023qnr,Alam:2021uuq,Gustafson:2022xdt,Lamm:2019bik,Zache:2023dko,Miranda-Riaza:2025fus}. Among these bases, the \emph{mixed basis} has emerged as one that can faithfully represent the gauge field across a wide array of different gauge couplings by combining both continuous magnetic variables and discrete electric variables to describe the wavefunction. 

This work studies the dynamics of non-stabilizerness in non-abelian LGTs by considering pure gauge SU(2) in the mixed basis. The smallest non-trivial system in this formulation is that of two interacting plaquettes, whose low-energy spectrum is described by smooth functions of two magnetic variables and a single electric variable. Central to this work is the development of a novel approach for mitigating errors in bitstring distributions, called Bit String Decoherence Renormalization (BSDR). This approach takes advantage of the fact that the amplitudes of such smooth wavefunctions admit a hierarchical expansion in terms of operators whose errors can be efficiently mitigated. Similar hierarchies have been utilized in the context of state preparation~\cite{welch2014efficient,zylberman2024efficient,Li:2024lrl} and error correction~\cite{klco2021hierarchical} for a variety of different quantum applications, like options pricing, machine learning, and solving differential equations~\cite{rebentrost2018quantum,gonzalez2024efficient}. This mitigation method, in conjunction with randomized measurement protocols, can additionally allow for the extraction of a number of non-trivial state properties beyond just the non-stabilizerness, like entanglement and the Inverse Participation Ratio~\cite{alam2026onset}.

This work is organized as follows. Section~\ref{sec:two_plaquette_section} introduces the Hamiltonian for this system; a detailed exposition of the encoding onto a quantum device is given in Ref.~\cite{Froland:2025bqf}. Section~\ref{sec:magic_numerics} introduces the relevant measure of non-stabilizerness and numerically examines its properties in both low-energy eigenstates as well as the dynamical setting of a quantum quench. Section~\ref{sec:walsh_qem} first presents the measurement protocol used to extract non-stabilizerness from the quantum device and introduces an improved estimator that significantly increases the efficiency of the calculation. Then, the novel mitigation technique is developed and its range of applicability is reported. Section~\ref{sec:quantum_results} reports the non-stabilizerness measurements on an IBM quantum computer, \textbf{ibm\_marrakesh}. Finally, Section~\ref{sec:discussion} offers concluding remarks and outlines directions for future work.

%%% %%% %%% %%% %%%
\section{SU(2) Lattice Gauge Theory on Two Plaquettes}\label{sec:two_plaquette_section}
This section provides a concise overview of constructing the fully gauge-fixed SU(2) lattice gauge theory Hamiltonian, as well as the explicit form of the digitized Hamiltonian for the two-plaquette system with open boundary conditions. Further details can be found in Refs~\cite{DAndrea:2023qnr, Grabowska:2024emw,Froland:2025bqf}.

The fully gauged-fixed Hamiltonian is constructed by applying two levels of gauge-fixing to the Kogut-Susskind (KS) Hamiltonian~\cite{PhysRevD.11.395}. The KS Hamiltonian is constructed out of electric field operators, $\hat{E}^b_{\ell L}, \hat{E}^b_{\ell R}$, defined on the links of a lattice, and magnetic plaquette operators, $\hat{P}_p$, constructed out of gauge link operators. The Hamiltonian itself is given by
\begin{align}
\hat H_\text{KS} = \frac{g^2}{2a}\sum_{\ell \in \text{links}}\hat E^2_\ell+ \frac{1}{2g^2a}\sum_{p \in \text{plaq.}}\Tr\left[2 - \hat P^{\phantom{\dagger}}_p-\hat P^\dagger_p\right]
\label{eq:HamGen}
\end{align}
where $g$ is the gauge coupling and $a$ is the lattice spacing, which is typically set to unity. Note that
\begin{align}
E^2_{\ell}\equiv E_{\ell L}^b E_{\ell L}^b =E_{\ell R}^b E_{\ell R}^b
\end{align}
where the index $b$ labels that color indices, which are summed over. The first level of gauge-fixing is carried out by applying maximal-tree gauge-fixing~\cite{PhysRevD.15.1128}, which systematically uses residual gauge transformations to set various gauge links to the identity. The gauge links that are set to the identity are non-dynamical and are called `tree' links; the remaining ones are dynamical and are conventionally called `physical'. This enforces all local Gauss law constraints, removing all local gauge redundancies. A second level of gauge-fixing is required to enforce the global Gauss-law, \ie to restrict the Hilbert space of the Hamiltonian to only span a specific total gauge charge sector. Typically the physical Hilbert space is chosen to be the total gauge singlet sector. This final layer of gauge-fixing is done by carrying out a change of variables on the group-element variables that define the Hilbert space of each (physical) link. This process has been completed for SU(2) in 2+1 and 3+1 dimensions in Refs.~\cite{DAndrea:2023qnr, Grabowska:2024emw}. 

\begin{figure*}[t]
    \centering
    \includegraphics[scale=.65, trim = 0 100 0 100]{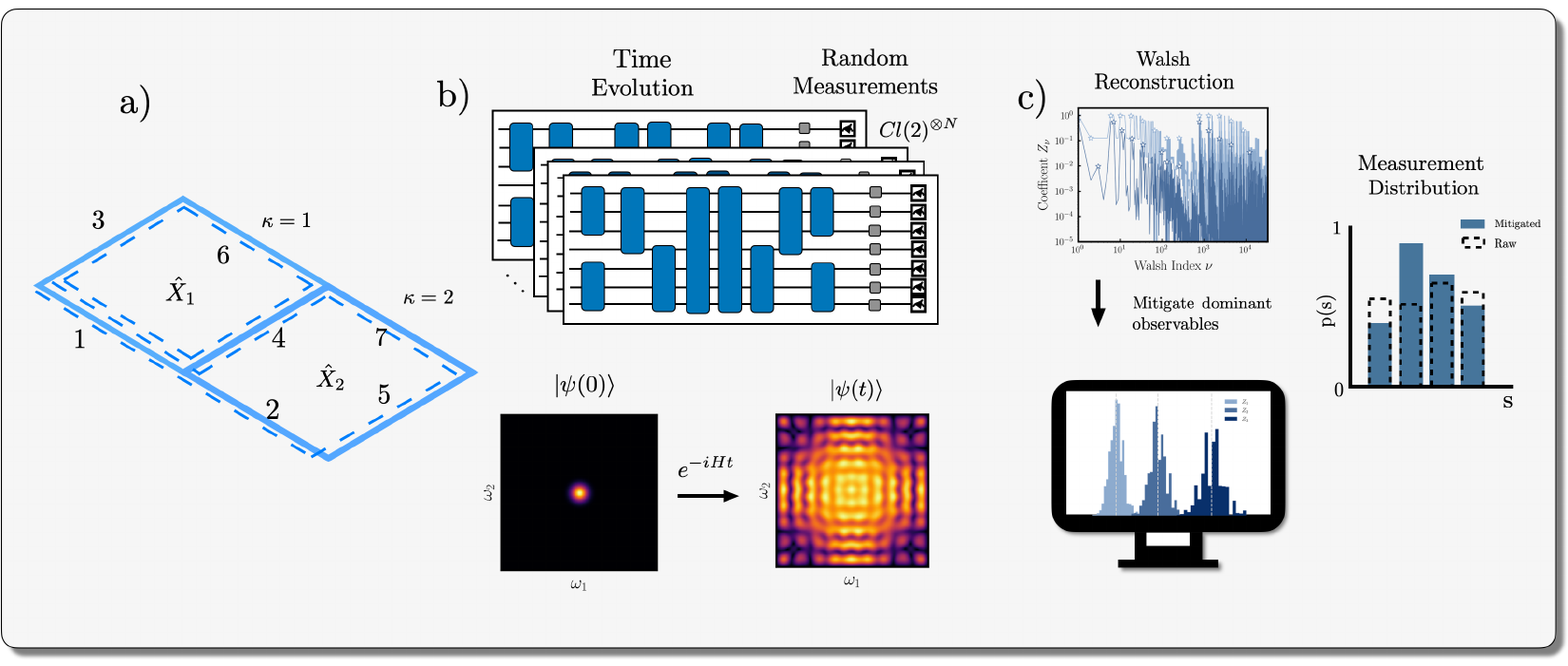}
    \caption{\textit{Digital Quantum Simulation of Two Plaquettes} \textbf{a)} The two-plaquette system with open boundary conditions is gauge-fixed using max tree gauge fixing. The system contains seven links, numbered $1$ through $7$, with the physical links denoted as $\kappa=1,2$. The magnetic degrees of freedom $\hat{X}_{1},\hat{X}_2$ are associated with each plaquette. The dashed lines that run around each plaquette denote the loops used in the max tree gauge-fixing procedure. \textbf{b)} The digitized Hamiltonian is compiled into digital quantum gates via Trotterization that approximate $e^{-iHt}$. Shown below the circuit diagram is an example of initial state amplitudes given by a smooth gaussian profile in $\omega_1,\omega_2$ and the final amplitudes after time evolution by the circuit. To measure the non-stabilizerness, many copies of the time evolved state are prepared and then a set of single qubit Clifford gates are sampled from the ensemble $Cl(2)^{\otimes n}$ and appended to the end of each time-evolved state. These copies are subsequently measured and the noisy bitstring distributions are recorded. \textbf{c)} From the set of random measurements, error mitigation can be applied to the expectation values of different Pauli Z strings $\langle Z_k\rangle$ to estimate the noiseless values. These mitigated expectation values can be used to reconstruct the error-mitigated bitstring distribution p(s), which can then be used to calculate non-linear functions of the state, like the Stabilizer R\'enyi Entropy or the purity.}
    \label{fig:intro_fig}
\end{figure*}

For two plaquettes, shown in Fig.~\SubFigRef{fig:intro_fig}{a}), the fully gauge-fixed Hamiltonian only depends on three variables, $\omega_{1}, \omega_2$ and $\theta$, where $\omega_i= [0, 2\pi)$ and $\theta = [0, \pi)$. The electric component of the Hamiltonian is given by~\cite{Froland:2025bqf}
\begin{align}
\label{eq:HEGE}
\hat{H}^{(E)}&=-\frac{g^2}{2}\left[\left\{4\left(\pdv[2]{}{\omega_1}+ \cot \frac{\omega_1}{2}\pdv{}{\omega_1}\right)-2 \csc^2\frac{\omega_1}{2}\CN\right.\right.\nonumber \\
&+\left.\left( \cot \frac{\omega_2}{2}\pdv{}{\omega_1}\right)\left(\sin \theta\pdv{}{\theta}\right)+\left(\omega_1 \leftrightarrow \omega_2\right)\right\}  \nonumber \\
&+ \left(\frac{1}{2}\cot \frac{\omega_1}{2}\cot \frac{\omega_2}{2}\right)\left(\sin \theta\pdv{}{\theta}\right)\nonumber \\
&-\left.\left(\frac{1}{2}\cos \theta \cot\frac{\omega_1}{2}\cot\frac{\omega_2}{2}-\frac{1}{2}\right)\CN\right]
\end{align}
where $\CN$ is the differential operator,
\begin{align}
\label{eq:LegDifEq}
\CN = - \pdv[2]{}{\theta}- \cot \theta \pdv{}{\theta} \, .
\end{align}
The magnetic component of the Hamiltonian is given by
\begin{equation}\label{HB}
    \hat{H}^{(B)} = \frac{2}{g^2}\left(2-\cos\frac{\omega_1}{2}-\cos\frac{\omega_2}{2}\right) \, ;
\end{equation}
note that the magnetic component of the Hamiltonian is diagonal in the group-element basis, as one would expect.

To transform from the group-element basis to the mixed basis, the continuous angular variable $\theta$ is changed into a discrete quantum number, $\ell$, via
\begin{align}
\braket{\theta}{\ell} = P_{\ell}(\theta)
\end{align}
where $P_{\ell}(\theta)$ are the (rescaled) Legendre polynomials of degree $\ell$.\footnote{The conventional normalization of the Legendre polynomials is 
\begin{align}
\int_0^\theta d(\cos \theta)\,P_{\ell'}(\theta)P_{\ell}(\theta) = \frac{2}{2\ell+1}\delta_{\ell'\ell}
\end{align}
but due to the requirement that the basis states that span a Hilbert space be orthonormal, the Legendre polynomials in this work are rescaled.} Note that here $\ell$ is a discrete quantum number whereas in Eq.~\eqref{eq:HamGen}, $\ell$ labels the links of the lattice. Legendre polynomials are used to span the Hilbert space because they are the eigenfunctions of the operator $\CN$,  
\begin{align}
    \CN P_{\ell}(\theta) = \ell(\ell+1) P_{\ell}(\theta) \, .
\end{align}
In the mixed basis, the amplitudes of any wavefunction are defined by
\begin{align}
\label{eq:RescalePsi}
\braket{\omega_1\, \omega_2\, \ell}{\psi} = \frac{u_{\ell}(\omega_1, \omega_2)}{4\sin\frac{\omega_1}{2}\sin\frac{\omega_2}{2}}
\end{align}
where $u_{\ell}$ is the regular part of the wavefunction and the factor in the denominator is introduced for computational simplicity.

In order to implement this Hamiltonian on a quantum device, the gauge field degrees of freedom have to be mapped into a register of qubits. This requires `digitizing' the continuous Hilbert space such that the physically-relevant features of the wavefunctions are faithfully and accurately represented. This is done in detail in Ref.~\cite{Froland:2025bqf} and only the relevant results are summarized here. The continuous degrees of freedom, $\omega_{1,2}$ are embedded into a register of $2 n_q$ qubits, with the resultant Hilbert space having dimension $2^{2n_q}$. This is done by equipartitioning the range $[0, \omega_\text{max})$ using $2^{n_q}$ points. The discretized variables are denoted $\omega_i^{(n)}$, for $i = 1,2$ and take on the values
\begin{align}\label{eq:grid_points}
\omega_i^{(n)} &= \delta_{\omega}\left(n+1/2\right) \qquad n \in \left\{0, 1, \dots, 2^{n_q}-1\right\} \nonumber \\
\delta \omega &= \omega_\text{max}/2^{n_q} \,\,.
\end{align}
Note that the value of $\omega_\text{max}$ depends on the value of the gauge coupling and the number of qubits,
\begin{align}\label{eq:optimal_w}
\omega_{\text{max}} = \text{min}\left(g(2^{n_q}-1)2^{(3-2n_q)/4}\sqrt{\pi},2\pi\right) \,\,.
\end{align}
The quantum number $\ell$ is naturally discrete and implemented onto $n_{\ell}$ qubits. Therefore the total size of the simulation Hilbert space is $2^{2n_q+n_{\ell}}$.

This discrete basis is used to map the infinite-dimensional Hamiltonian to a finite matrix representation. Operators that are functions of $\omega_i$ are diagonal and therefore the digitized operators are constructed by decomposing the $\omega_i$ variables into Pauli Z strings. The first- and second-order derivative operators are not diagonal in this space and care must be taken to construct their digitized representation. The second-order derivative is implemented using the exact lattice Laplacian with Dirichlet boundary conditions. This is given by
\begin{align}\label{eq:exact_laplacian}
\pdv[2]{}{\omega} &\mapsto \text{DST}^{-1}_{II}(-\hat{k}^2)\text{DST}_{II}\,\,
\end{align}
where $\hat k$ is a diagonal matrix with eigenvalues $\pi k/\omega_{max}$ with $k \in \{1, 2, \dots, 2^{n_q}\}$ and $\text{DST}_{II}$ is a variation of the Discrete Sine Transform (DST), 
\begin{align}
    (\text{DST}_{II})_{kn} = 
    \begin{cases}
    \sqrt{\frac{2}{N}}\sin\frac{\pi(n+\frac{1}{2})(k+1)}{N}, & \text{for } k \neq N-1\\
    \sqrt{\frac{1}{N}}\sin\pi(n+\frac{1}{2}),  & \text{for } k = N-1
    \end{cases} \,\,.
\end{align}
The specific form of $\text{DST}_{II}$ is chosen such that the field space has the appropriate boundary conditions. The first-order derivative is implemented using the finite central difference,
\begin{align}\label{eq:finite_central_diff}
    \left(\pdv{}{\omega}\right)_{kn} = 
    \begin{cases}
    \frac{1}{2\delta\omega}, & \text{for } k-n=1,\\
    -\frac{1}{2\delta\omega}, & \text{for } k-n=-1,\\
    0 & \text{else}
    \end{cases} \,\, .
\end{align}
Ref.~\cite{Froland:2025bqf} demonstrated that five qubits (two for each $\omega$ register and one for the $\ell$ register) were sufficient for estimating the expectation value of certain observables, such as the energy, with per-mille precision for a variety of gauge couplings. 

\section{Non-Stabilizerness of Two Plaquettes}\label{sec:magic_numerics}
Having mapped the continuous Hilbert space of the gauge field into a finite-dimensional quantum system, its quantum complexity can be quantified by the non-stabilizerness. On a system of $n$ qubits the Pauli group\footnote{To form a group, $\mathcal{P}$ must also include the possible phases $\pm 1,\pm i$. It is standard convention in the literature to refer to the set of all $n$ qubit Pauli operators without phases as the Pauli group.} is defined as $\mathcal{P}=\{\otimes_i^n\sigma_i\;|\; \sigma_i\in I,X,Y,Z\}$. The $n$ qubit Clifford group $Cl(n)$ is defined as the group of unitaries that normalize $\mathcal{P}$, and the stabilizer states are the set of states that are fixed by the action of $Cl(n)$. Computations involving only gates from $Cl(n)$ acting on stabilizer states can be performed efficiently classically~\cite{aaronson2004improved}. The non-stabilizerness quantifies the distance of a state from the set of stabilizer states, and as such captures one notion of the quantum complexity of a state.

A measure of non-stabilizerness~\cite{leone2022stabilizer} that is computable for larger systems of qubits is the Stabilizer R\'enyi Entropy (SRE). A pure state $\rho=\ket{\psi}\bra{\psi}$ is decomposed as
\begin{equation}
    \rho = \frac{1}{d}\sum_{P\in\mathcal{P}}c_PP,\;\;\;c_P=\text{tr}\rho P
\end{equation}
where $d=2^n$. The purity of the state enforces the condition $\sum_P|c_P|^2/d=1$, implying that the set of squared coefficients $|c_p|^2/d$ form a probability distribution. The entropies of this distribution are the $\alpha$-Stabilizer R\'enyi Entropies and are defined by
\begin{equation}\label{eq:sre_defition}
    \mathcal{M}_{\alpha}=\frac{1}{1-\alpha}\log\frac{1}{d}\sum_P|c_P|^{2\alpha}
\end{equation}
This quantity vanishes for stabilizer states and is non-increasing under application of non-Clifford unitaries for $\alpha\geq 2$, making it a good measure of non-stabilizerness from a resource theory perspective~\cite{leone2024stabilizer}. Consequently, this works uses the specific case of $\mathcal{M}_2$. A system can be said to have a large amount of non-stabilizerness if, for example, $\mathcal{M}_2$ grows extensively with system size. For mixed states the definition of the SRE must be modified since the set of Pauli expectations no longer forms a distribution. In this case, these expectations must be normalized by the purity, which is expressed as
\begin{equation}
    S_2 = \frac{1}{d}\sum_{P}|c_P|^2
\end{equation}
and allows the SRE to be defined for mixed states. This definition only reliably detects stabilizer states when the purification of the state is itself a stabilizer~\cite{trigueros2025nonstabilizerness}.

\subsection{Non-Stabilizerness of Low-Energy Eigenstates}
The low-energy spectrum of the digitized mixed-basis is comprised of smooth functions encoded into wavefunction amplitudes. Understanding the complexity of preparing such states will inevitably involve describing the stabilizer properties of such wavefunctions. Previous studies of ground state non-stabilizerness in the electric basis~\cite{Santra:2025dsm} show markedly different behavior than in the mixed basis, emphasizing the importance of basis choice when studying the theory at different values of the gauge coupling. The electric basis diagonalizes the Hamiltonian in the strong coupling limit, and so the eigenstates in this limit are trivially stabilizers. The mixed basis serves as an intermediate choice between the electric and magnetic representations and the eigenstates in the mixed basis exhibit non-vanishing non-stabilizerness at all values of the gauge coupling.

\begin{figure*}[t]
    \centering
    \includegraphics[scale=.6, trim = 80 130 40 150]{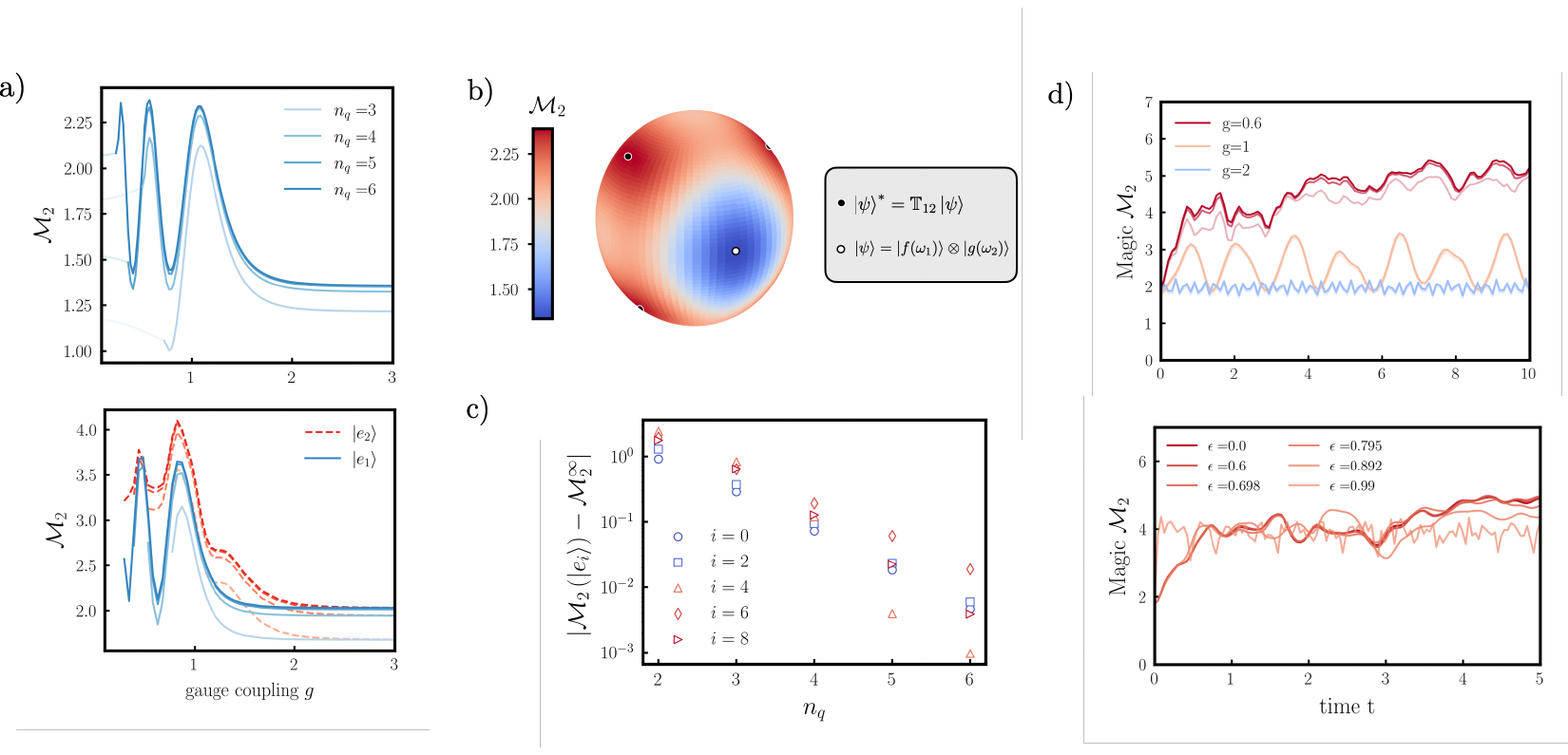}
    \caption{\textit{Non-Stabilizerness of Low-Energy Eigenstates} \textbf{a)} Non-stabilizerness of ground state $\ket{e_0}$ (top) and first two excited eigenstates $\ket{e_1},\ket{e_2}$ (bottom). For large gauge coupling, the non-stabilizerness plateaus to a constant value for $\ket{e_1},\ket{e_2}$, which are degenerate in this limit. For a fixed resolution, once a certain coupling is reached, the $\omega_{\text{max}}$ truncation begins to decrease from $2\pi$, causing the effective support of the wavefunction over $\omega_1-\omega_2$ to change at a different rate with $g$, which is signified by the transparent lines (these lines are not shown for the bottom plot). \textbf{b)} Non-stabilizerness in degenerate subspace $\mathcal{S}_{12}$. The states that have the minimum non-stabilizerness in this subspace are separable, while the states that have the maximum non-stabilizerness can be characterized by their properties under $\omega_1-\omega_2$ exchange. \textbf{c)} Convergence of non-stabilizerness with increasing digitization resolution for first ten even eigenstates. \textbf{d)} Time evolution of $\mathcal{M}_2$ for different values of gauge coupling (top) and different levels of Pauli truncation $\epsilon$ in Hamiltonian (bottom). In the top plot, multiple digitization resolutions ranging from $n_q=3$ to $n_q=6$ are shown for each value of the coupling.}
    \label{fig:twop_magic}
\end{figure*}

Figure ~\SubFigRef{fig:twop_magic}{a}) (top) shows $\mathcal{M}_2$ for the ground state as a function of the gauge coupling. For large values of the coupling, the non-stabilizerness approaches the value calculated using the strong-coupling ground state wavefunction. In this limit, the wavefunction has the functional form $u_0(\omega_1,\omega_2)=\sin\frac{\omega_1}{2}\sin\frac{\omega_2}{2}$ and the non-stabilizerness is $\mathcal{M}_2\sim1.4$. As the gauge coupling is decreased, the support of the wavefunction, \ie where it is not exponentially small, starts to decrease and becomes concentrated around $\omega_1,\omega_2$ close to zero. At intermediate values of the gauge coupling, $g\sim \mathcal{O}(1)$, the non-stabilizerness oscillates log-periodically, \ie with minima that occur regularly with respect to $\log g$. Once $g$ reaches the value such that $\omega_{\text{max}}$, as given by Eq.~(\ref{eq:optimal_w}), is less than $2\pi$, the range of $\omega$ decreases with the support of the wavefunction and qualitatively different behavior is observed.

To understand the origin of these oscillations, it is instructive to consider a simpler wavefunction that is constant for $\omega_1,\omega_2<C$ for $0\leq C\leq 2\pi$ and vanishing outside this region. When such a state is encoded into a system of $2n$ qubits, the cutoff $C$ will correspond to an integer between $0$ and $2^{n}-1$, as in Eq.~(\ref{eq:grid_points}). As shown in Appendix~\ref{app:magic_indicator}, the non-stabilizerness of such constant wavefunctions will have minima when the integer representation of $C$ coincides with a power of two. When the wavefunction varies over its range of support, as is the case with the ground state of the two-plaquette system, there will instead be a minimum near a power of two. As the support of the ground state wavefunction decreases with $g$, the observed log-periodic oscillatory behavior occurs.

The behavior of non-stabilizerness in the first and second excited eigenstates $\ket{e_1}$ and $\ket{e_2}$, shown in Fig.~\SubFigRef{fig:twop_magic}{a}) (bottom), closely mirrors that of the ground state, with saturation at large gauge coupling followed by subsequent oscillatory behavior as the gauge coupling decreases. The most striking feature is that the non-stabilizerness of both eigenstates approaches the same value in the strong coupling limit, owing to the degeneracy of $\ket{e_1}$ and $\ket{e_2}$ in the same limit. This degeneracy is related to both the separability of $H_E$ and the $\omega_1\leftrightarrow\omega_2$ swap symmetry implemented by the operator $\mathbb{T}_{12}=\frac{1}{d}\sum_i\sigma_{1,i}\otimes\sigma_{2,i}$, where $\{\sigma_{j,i}\}$ denotes the set of single qubit Pauli operators on the $i^{th}$ qubit of each of the $\omega_j$ registers. In the strong coupling limit, $\ket{e_1}$ and $\ket{e_2}$ form a two-dimensional subspace $\mathcal{S}_{12} = \text{span}(\ket{e_1},\ket{e_2})$ that is composed of states of the form $\ket{\psi(\vartheta,\varphi)}=\cos\frac{\vartheta}{2}\ket{e_1}+e^{i\varphi}\sin\frac{\vartheta}{2}\ket{e_2}$ and is parameterized by the two angles $(\vartheta,\varphi)$. In this subspace, $H_E$ is diagonal whereas $H_B$ is not. Treating $H_B$ as a perturbation for large but finite $g$ using degenerate perturbation theory, it is seen that the basis states of $\mathcal{S}_{12}$ that diagonalize this perturbation have the same non-stabilizerness.

This equivalence may naively suggest that from the perspective of Hamiltonian simulation, the best basis to choose is the one that is aware of the structure of the perturbation, namely the one that splits the degeneracy. However, considering the range of values of the non-stabilizerness over the entire subspace $\mathcal{S}_{12}$ reveals a different picture. Figure ~\SubFigRef{fig:twop_magic}{b}) shows the non-stabilizerness over $\mathcal{S}_{12}$ as a function of $\vartheta$ and $\varphi$. The states that maximize and minimize non-stabilizerness over $\mathcal{S}_{12}$ can be characterized by how they transform under $\mathbb{T}_{12}$. The basis is chosen such that the coordinate $(0,0)$ corresponds to the state that is symmetric under $\omega$ exchange while the coordinate $(\pi,0)$ corresponds to one that is anti-symmetric. The states that minimize the non-stabilizerness lie at the coordinates $(\pi/2,0)$ and $(\pi/2,\pi)$, corresponding to separable states of the form $\ket{\psi}=\ket{f(\omega_{1})}\otimes\ket{g(\omega_{2})}$ with $f(\omega)=\sin\frac{\omega}{2}$ and $g(\omega)=\sin \omega$. The states with the highest non-stabilizerness are defined by the coordinates $(m\pi/4,n\pi/2)$ with $m,n\in\{1,3\}$. These states can be defined by the property that exchange of $\omega_1,\omega_2$ is equivalent to complex conjugation of the amplitudes, \ie $\ket{\psi}^*=\mathbb{T}_{12}\ket{\psi}$. This demonstrates that the most advantageous basis for quantum simulation may not always be the one that respects the structure of the Hamiltonian, and instead is the one that minimizes complexity.

Finally, the convergence of the non-stabilizerness with increasing digitization resolution is studied. Low digitization resolutions have the effect of coarse-graining small-scale variations in the wavefunction amplitudes, lowering the perceived non-stabilizerness. As the resolution is increased, these variations can have a non-trivial contribution to the resource requirements, causing the non-stabilizerness to grow extensively with the number of qubits. On the other hand, for smooth wavefunctions whose frequency spectrum decays quickly, increasing the resolution does not noticeably change the overall structure of the state and the non-stabilizerness will plateau below the expectation for typical, Haar random, quantum states. For the purposes of simulating these states in the continuum limit, this suggests that it is only necessary to simulate the state up to a minimum resolution, aligning with the expectation that the mixed basis requires a low number of quantum resources for different values of $g$.

For low-energy states $\ket{e_i}$, the convergence to the infinite resolution limit ($n_q\rightarrow\infty$) is exponential. To understand how quickly this convergence happens, the non-stabilizerness for the first ten eigenstates is calculated for a fixed coupling of $g=1$. The non-stabilizerness at infinite resolution $\mathcal{M}^{\infty}_2$ is extracted by fitting its values at finite $n_q$ to the curve $\mathcal{M}^{n_q}_2(\ket{e_i})=ae^{-b n_q}+\mathcal{M}^{\infty}_2$, with the finite $n_q$ deviation being shown in Fig.~\SubFigRef{fig:twop_magic}{c}). While there is modest variance in the rates of convergence, all but one of the states studied reaches $10^{-2}$ precision with respect to the continuum value once $n_q=6$. This stands in contrast to the required digitization levels for extracting values for local observables such as the energy density, for which it has been shown that as few as $n_q=3$ qubits are required to get within per-mille precision of the continuum value~\cite{Froland:2025bqf}. This difference demonstrates that capturing the full complexity of a quantum state is not necessary for calculating the value of local observables in quantum simulations.

%%%   %%%   %%%
\subsection{Time Evolution and Truncation Effects}
The convergence of non-stabilizerness with digitization in the low-energy subspace becomes especially relevant when considering time evolution. This behavior is studied by considering the following quench-like scenario: The system is prepared in a superposition of low-lying eigenstates of the Hamiltonian at some fixed gauge coupling $g_0$ then evolved with the Hamiltonian at a different coupling $g$. The exact superposition is not important, only that the states in it are sufficiently far from the middle of the spectrum so that over the ranges of times considered the time-evolved state does not mix with energy bands that contain a number of states that grow extensively with system size. For concreteness, the initial state is an equal superposition of the first three eigenstates, $\ket{\psi(0)}=(\ket{e_0}+\ket{e_1}+\ket{e_2})/\sqrt{3}$, which has energy on the order of the mass gap of the system.

Fig.~\SubFigRef{fig:twop_magic}{d}) (top) shows the time-evolved non-stabilizerness for $g_0=10$, very close to the strong-coupling limit, for $g=\{0.6,1,2\}$. For $g=2$ the time evolution does not rapidly mix eigenstates of the $g_0$ Hamiltonian and therefore the non-stabilizerness exhibits small oscillations between the initial states. As $g$ is decreased, the time evolution induces mixing between the initial state and other eigenstates in increasingly higher energy bands, causing the oscillations to grow and then subsequently diminish as the dynamics become dominated by ramp-like behavior. The same type of growth is observed when initializing the system in eigenstates corresponding to small $g$ and then time evolving at values of larger $g$. This behavior can be understood by recognizing that the value of the gauge coupling changes the eigenbasis of the Hamiltonian and therefore time-evolving the eigenstate of one Hamiltonian with another Hamiltonian has the effect of introducing more components into the functional decomposition of the wavefunction. The number of components that contribute non-trivially will depend on how closely the two eigenbases overlap. This suggests, that for wavefunctions representing smooth functions, a greater number of components will give a higher amount of non-stabilizerness.

When simulating real-time dynamics on a quantum device, it is necessary to also consider how approximations to the time evolution operator modify the complexity of the prepared state. The circuits in this work are constructed by decomposing the Hamiltonian into a sum of Pauli strings and dropping strings that have sufficiently small prefactors from the decomposition. When viewed from the perspective of Trotterization, this has the effect of removing many small $R_z$ rotations (up to Cliffords) and so it is expected that as long as the prefactor is small enough the complexity of the state will not be affected. Figure ~\SubFigRef{fig:twop_magic}{d}) (bottom) shows the time evolution of $\ket{\psi(0)}$, where $\epsilon$ percent of Pauli string with the smallest coefficients are removed from the Hamiltonian. Surprisingly, as many as $\epsilon\approx 0.6$ of the Paulis, corresponding to $60\%$ of the total operator content of the Hamiltonian, can be removed with no visual change to the non-stabilizerness, allowing for the complexity dynamics to be studied on the quantum device for deeper circuit depths.

\section{Walsh Based Error Mitigation}\label{sec:walsh_qem}
This section outlines the techniques used for measuring the non-stabilizerness on a quantum device. The distribution of measured bit strings is the primary observable used to calculate $\mathcal{M}_2$. Obtaining accurate estimates from quantum hardware necessitates the development of techniques that allow one to account for the effects of noise present on a device. Since the non-stabilizerness depends on the entire bitstring distribution, without \emph{a priori} knowledge of this distribution, naive scaling arguments would imply a mitigation overhead that grows proportionally to the number of bitstrings, \ie exponentially with system size. An approach tailored for mitigating the noise in bit string distributions, that in certain cases ameliorates the dependence on system size, is presented. It is shown that for wavefunctions that are the digitizations of smooth functions, e.g. eigenstates of the two-plaquette system, it is sufficient to mitigate a constant number of observables.

\subsection{Random Measurements and Improved Estimator}
Measuring the non-stabilizerness of a state on a quantum device is made possible through the use of randomized measurements~\cite{oliviero2022measuring}. The protocol begins by randomly sampling $N_U$ single-qubit Clifford unitaries, forming an ensemble of circuits that are averaged over. Once the desired state is prepared, the random Clifford rotations are applied to each qubit, measured $N_M$ times and the outcome bitstrings are recorded. This set up is illustrated in Fig.~\SubFigRef{fig:intro_fig}{b}). The collection of $N_M$ outcomes, per random Clifford, gives an estimate for a probability distribution in the computational basis $\ps$, where the dependence on the Clifford unitary has been suppressed. For the measured distribution, the following estimators are computed
\begin{align}\label{eq:magic_estimators}
    O_j^{(2)} &= \sum_{\mbs_1,\mbs_2}2^{-w(\mbs_1\oplus\mbs_2)}\text{p}(\mbs_1)\text{p}(\mbs_2)\\
    O_j^{(4)} &= \sum_{\mbs_1,\mbs_2,\mbs_3,\mbs_4}2^{-w(\mbs_1\oplus\mbs_2\oplus\mbs_3\oplus\mbs_4)}\nonumber\\
    &\hspace{60pt}\text{p}(\mbs_1)\text{p}(\mbs_2)\text{p}(\mbs_3)\text{p}(\mbs_4)
\end{align}
where $w(\mbs)$ denotes the Hamming weight of the bitstring $\mbs$ and the subscript $j\in\{1,\dots,N_U\}$ indexes the circuit in the measurement ensemble. Averaging $O^{(2)}_j,O^{(4)}_j$ over the ensemble of random Cliffords provides experimental estimates of the Stabilizer R\'enyi Entropy $\mathcal{M}_2$ and the purity $S_2$, which are given by 
\begin{align}
\mathcal{M}_2 &= -\log X^{(4)} + \log dX^{(2)} - \log d \nonumber \\
S_2 &= d X^{(2)}
\end{align}
in the limit $N_U\rightarrow\infty$, where $X^{(\ell)}=\frac{1}{N_U}\sum_{j=1}^{N_U}O^{(\ell)}_j$.

The expressions for the estimators $O_j^{(2)}$  and $O_j^{(4)}$ involve summing over $16^n$ terms and so the classical post-processing cost of the calculation can be prohibitive for large $n$. It is beneficial to introduce an improved estimator, which is simply a re-writing of Eq.~(\ref{eq:magic_estimators}). The derivation fundamentally relies on the Walsh-Hadamard (WH) transform of a Boolean function $f(s)$, denoted $\hat{f}(\mbk)=\sum_{\mbs}f(\mbs)(-1)^{\mbk\cdot \mbs}$. For notational convenience the WH transform will be denoted as $c_\mbk=\hat{f}(\mbk)$ and referred to as the \emph{Walsh coefficients}. While the use of the WH transform has recently been recognized to give advantages in similar contexts~\cite{Huang:2025cbk, Xiao:2026ptn,Sierant:2026jru}, the fact that the estimators $O^{(\ell)}_j$ are written in terms of bitstrings allows the equations to be written in closed form. Specifically for $\ell = 2, 4$, the estimators are given by
\begin{equation}\label{eq:decoupled_magic_estimators}
    O^{(\ell)}_j = \frac{1}{d^2}\sum_{\mbk}3^{w(\mbk)}\hat{\text{p}}_j^{\ell}(\mbk)\, ;
\end{equation}
details of this derivation are in Appendix~\ref{app:estimator_results}. The key observation is that while the original expression for $\tilde{\mathcal{M}}_2$ involved summing over $16^{n}$ different terms, the improved expression only sums over $2^n$. Implementing the WH transform requires $\mathcal{O}(n2^n)$ operations, but it is also possible to put single-qubit Hadamards after the sequence of random Cliffords and then make computational basis measurements. This reduces post-processing to strictly $\mathcal{O}(2^n)$ at the potential cost of extra hardware error. Single-qubit error rates are generally an order of magnitude below that of two-qubit gates and so this is not expected to be a large contribution to the error budget on current-day NISQ-devices. This reduction in complexity enables the estimators to be easily evaluated on a laptop for the same system sizes that are accessible via exact diagonalization ($n\sim24$).

\subsection{Error Mitigation Technique}
Error mitigation refers to a host of techniques that aim to reduce the effects of device noise on measured quantities, often involving application of compensating factors in post-processing. Many powerful techniques have already been developed that correct noisy expectation values of operators ~\cite{cai2023quantum,endo2018practical,wallman2016noise,van2022model}. However, the observables $O^{(2),(4)}$ are functions of $\ps$ and require getting mitigated estimates of the bitstrings themselves. Different strategies for accomplishing this have been developed in different contexts, including matrix inversion ~\cite{Nation:2021kye}, maximum likelihood estimation~\cite{Lee:2026xfm,Chandramouli:2025rfx,Baron:2024kdb}, and Monte-Carlo sampling~\cite{Liu:2025glx}. This work introduces a novel approach to error mitigation called Bit String Decoherence Renormalization (BSDR), aimed at mitigating the bitstring distributions $\ps$ relevant for calculating quantities like non-stabilizerness or R\'enyi entropies.

A quantum state $\ket{\psi}$ written in the computational basis is given by $\ket{\psi}=\sum_{\mbs}e^{i\phi_{\mbs}}\sqrt{\ps}\ket{\mbs}$. Similarly, projectors onto the computational basis states can be written as 
\begin{align}
\ketbra{\mbs}{\mbs}=\sum_{\mbk}(-1)^{\mbk\cdot\mbs}Z_\mbk
\end{align}
where $Z_\mbk$ denotes a string of Pauli $Z$ operators and $k$ is a length $n$ bitstring. Here, $Z_k$ means there is an identity on qubit $j$ if the $j^{th}$ entry of $k$ is $0$ and $Z$ where the $j^{th}$ entry is $1$. The desired bitstring distribution  can therefore be formulated as an expectation value 
\begin{align}
\ps=\text{Tr}\rho\ketbra{\mbs}{\mbs}
\end{align}
with $\rho=\ketbra{\psi}{\psi}$. It then follows straightforwardly that the bitstring distribution $\ps$ is just the WH transform of the Walsh coefficients $c_\mbk$, which coincide with the set of expectation values for $Z$ strings. This suggests that to mitigate the distribution $\ps$, one should mitigate the $Z$ expectation values and then reconstruct $\ps$ from the set of mitigated observables. A secondary implication of the equivalence between $\ps$ and $c_\mbk$ is that the estimators Eq.~(\ref{eq:decoupled_magic_estimators}) for non-stabilizerness and purity can be expressed directly in terms of the Clifford averaged $Z$ expectation values. Using this approach, the noisy estimate for the bitstring probability is given by
\begin{equation}
    \tilde{\text{p}}(\mbs) = \sum_\mbk (-1)^{\mbk\cdot\mbs}\tilde{c}_\mbk
\end{equation}
where the tilde denotes the mitigated estimate for both $\ps$ and $c_\mbk$. Defining the bias of the estimate to be $\text{Bias(\text{p}(s))}=\tilde{\text{p}}(\mbs)-\ps$, then
$\text{Bias}(\tilde{\text{p}}(\mbs)) = \sum_\mbk (-1)^{\mbk\cdot\mbs}\text{ Bias}(c_\mbk)$, which implies that when the mitigation method for the observables is unbiased, the estimator $\tilde{\text{p}}(\mbs)$ is also unbiased.

To understand why inferring $\ps$ from mitigated observables is advantageous, it is important to note that the noise affecting hardware can generally be modeled as Pauli noise
\begin{equation}
    \mathcal{E}(\rho) = \sum_{P}a_P P\rho P
\end{equation}
where the sum runs over all Pauli strings and $0\leq a_P \leq 1\;\;\forall\, P$. The effect on an expectation value is simply rescaling by a coefficient $r$, \ie $\text{Tr}\mathcal{E}(\rho)O=r\text{Tr}\rho O$, with $r$ defined as
\begin{equation}
    r=\sum_{P\in O^+}a_p-\sum_{P\in O^-}a_p
\end{equation}
where $O^{+/-}$ are the set of Pauli strings that commute/anti-commute with $O$. While Pauli noise acts in a predictable way for Pauli string observables, the effect on bitstrings is generally more complicated and not just a rescaling. Observable mitigation methods like Zero-Noise Extrapolation (ZNE) and Operator Decoherence Renormalization (ODR)~\cite{Urbanek:2021oej,ARahman:2022tkr,Farrell:2023fgd,Farrell:2024fit} take advantage of this structure, assuming that the noise strength is weak enough so that $r$ can be canceled. While there is likely a regime of noise strength for which a simple relation between the bitstring probabilities and noise strength holds, the fact that $\ps$ is a combination of many Pauli strings obfuscates the point at which this limit becomes appropriate. Therefore, applying the wealth of methods for mitigating observables $O$ and using these mitigated values to infer the bitstring distribution is found to be much more effective.

While this strategy is flexible enough to be used in conjunction with any other approach for mitigating observables, it becomes particularly powerful when paired with a variant of decoherence renormalization, ODR. The values of $p_\mbk$ are `learned' by running a \textit{mitigation} circuit alongside the circuit of interest, referred to as the \textit{physics} circuit. A mitigation circuit is one that can be classically efficiently simulated so that it has known output, but is structurally similar to the physics circuit. The mitigation circuit is designed so that its noise profile is a good approximation of the physics circuit. This work uses \textit{self-mitigating} mitigation circuits, a style of mitigation circuit that takes advantage of the symmetrized structure of a second-order Trotter step, $U^{\dagger}(-t/2)U(t/2)$ where $U$ is a first-order step, to implement time evolution to time $t$. When the signs of the angles in the second part of the circuit are flipped, the evolution due to the first part is reversed and so the overall circuit simply implements the identity. Using this mitigation circuit, the noisy expectation value of some observable of interest is compared to its true value, allowing the noise parameter to be estimated as $p_\mbk=\left( c_\mbk\right)_{\text{mit}}^{\text{true}}/\left( c_\mbk\right)_{\text{mit}}^{\text{meas}}$, where \textit{true}/\textit{meas} denote the exact value and the measured value respectively.

\subsection{Walsh Based Reconstruction}
It is undesirable to require mitigating every $c_\mbk$, as their number scales exponentially with the number of qubits. However, for states that encode smooth functions such as the low-energy states of the two-plaquette system, the Walsh coefficients admit a hierarchy that allows for efficient reconstruction. This idea has been used before in the context of compression for digital signals. To understand how this hierarchy appears, it is necessary to define some notation. The \emph{sequency} of a bitstring is the digital analog of frequency, defined as the number of sign changes in the row of the
n-qubit Hadamard matrix $H_2\otimes\dots\otimes H_2$, with $H_2 = (X+Z) / \sqrt{2}$ the single-qubit Hadamard, indexed by that bitstring under natural (binary) ordering. The sequency ordering is then obtained by re-indexing rows according to the decimal value of their bit-reversed Gray code: for each bitstring $k$, one computes its Gray code representation, reverses the order of its bits, \ie reflects the bitstring left-to-right, and ranks rows by the resulting decimal values in ascending order. The $\nu^{\text{th}}$ sequency band is defined by the collection of sequency ordered bitstrings that lie in the range $2^{\nu-1}$ to $2^{\nu}-1$. For multivariate functions defined on a $D$-dimensional interval ($D=2$ for the two-plaquette eigenstates), each band is characterized by a tuple of band coefficients. As shown explicitly in Appendix.~\ref{app:reconstruction_bounds}, for sufficiently smooth functions the coefficients in each band can generally be bounded as 
\begin{equation}\label{eq:walsh_hierarch}
    c_{\nu} \sim \prod_{i=0}^D \frac{1}{2^{\nu_i}}
\end{equation}
which gives a prescription for how reconstruction should proceed.
\begin{figure*}[t]
    \centering
    \includegraphics[scale=.55, trim = 100 50 100 50]{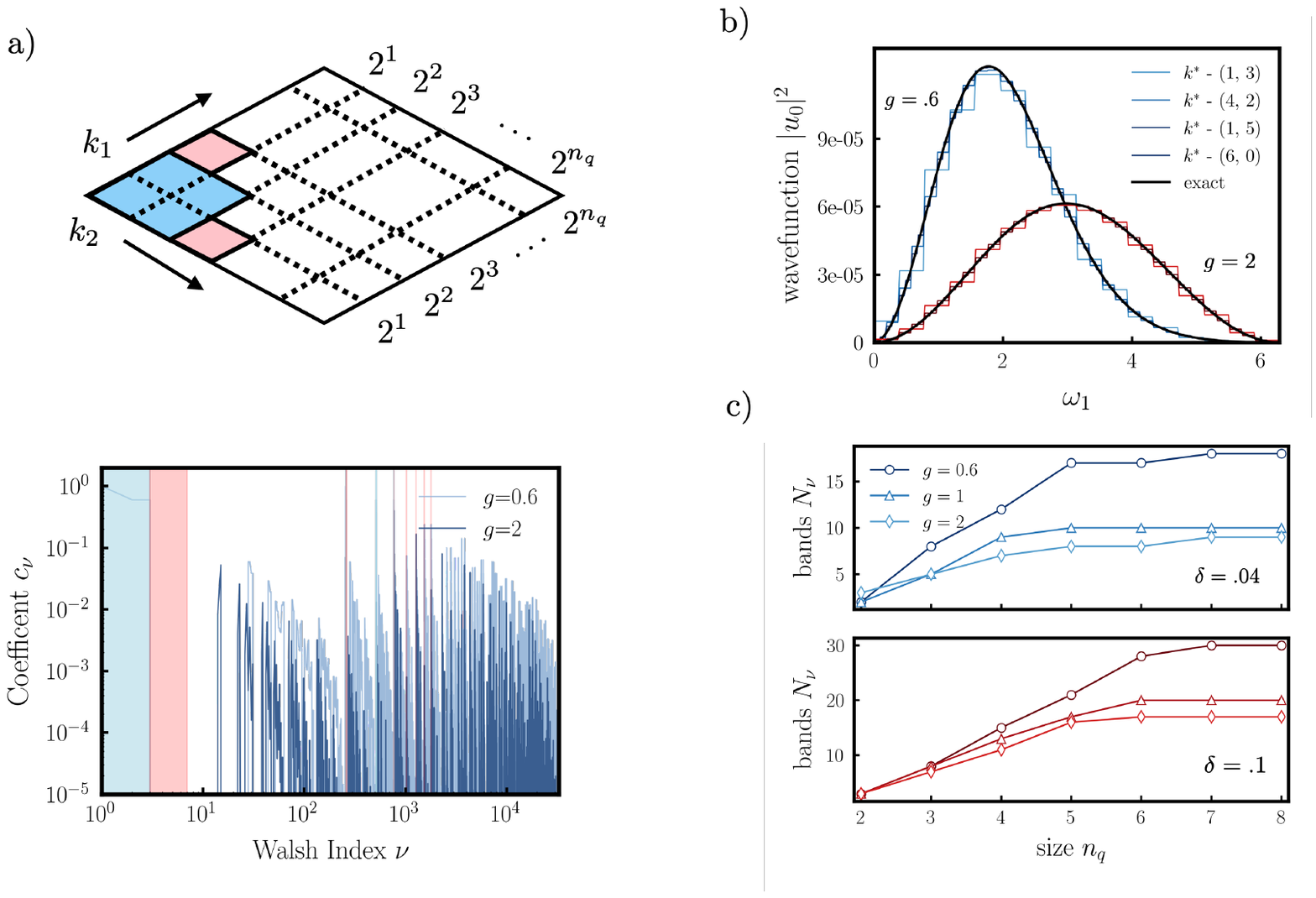}
    \caption{\textit{Walsh Hierarchy of Smooth Wavefunctions} \textbf{a)} Different Walsh bands for two-dimensional function of $\omega_1,\omega_2$ (top) with Walsh indices and associated coefficients (below). The lowest, most significant, bands are highlighted in blue, and the next most significant are shown in red. The strategy to reconstruct the wavefunction involves starting from the shells closest to the origin and working out in each direction. Once a shell is filled, the next shell starts being filled. \textbf{b)} Reconstructed wavefunction probability density for fixed $\omega_2$ at different Walsh cutoffs $k^*$. The solid black lines correspond to exact reconstruction where the cutoff is removed. Both a strong coupling wavefunction at $g=2$ and a weak coupling wavefunction at $g=0.6$ are shown. \textbf{c)} Number of Walsh bands needed to reconstruct distribution $\text{p}(\omega_1,\omega_2)$ up to fixed error $\delta=.1/.04$ (top/bottom) as function of $n_q$. The number of bands in each case increases linearly, corresponding to an exponential growth in the number of required Walsh coefficients until a plateau is reached. The different points in the plot denote the values of $g$ for which the ground state wavefunction was calculated.}
    \label{fig:reconstruction_fig}
\end{figure*}

Fig.~\SubFigRef{fig:reconstruction_fig}{a}) describes the strategy for reconstruction, taking the ground state of the two-plaquette system at couplings $g=0.6$ and $g=2$ as concrete examples. The bands with greater influence on the reconstruction will have the lowest band indices $(\nu_1,\nu_2)$, and so starting from the origin the index of each band is incremented individually, alternating between $\nu_1$ and $\nu_2$. The bottom figure explicitly shows the entire sequency spectra, with the bands containing the lowest, most dominant coefficients highlighted. It is actually unnecessary to calculate all of the coefficients within each band, as only including the most dominant coefficient generally gives the largest improvement. However, how these maximum coefficients are distributed within bands is non-trivial, and without further knowledge of the specific function, one must mitigate the entire band to account for its contribution. Fig.~\SubFigRef{fig:reconstruction_fig}{b}) shows the reconstruction quality at various cutoffs compared to the exact wavefunction for fixed $\omega_2$. For low cutoffs only the shape of the distribution is recovered, although as the cutoff is increased the profile of the reconstructed distribution becomes indistinguishable from the exact curve.

To measure the quality of the reconstruction, denote the reconstructed distribution up to sequency number $\mbk^*$ as $\psk{\mbk^*}$. The error between the exact distribution $\ps$ and $\psk{\mbk^*}$ is quantified by the Total Variational Distance (TVD), $\delta$, defined by
\begin{equation}
    \delta(\text{p}_1,\text{p}_2) = \frac{1}{2}\sum_\mbs |\text{p}_1(\mbs) - \text{p}_2(\mbs)|
\end{equation}
and which satisfies the properties of a distance over distributions. Assuming the hierarchy in Walsh coefficients satisfies the condition Eq.~\eqref{eq:walsh_hierarch}, it is shown in Appendix~\ref{app:reconstruction_bounds} that for a $D$-dimensional function the TVD is bounded as
\begin{equation}\label{eq:tvd_bound}
    \delta{}\leq \sqrt{\frac{DC^D}{6}\Delta}\sum_j2^{-m_j}
\end{equation}
where $m_j=\lfloor{\log_2 k_j^*}\rfloor$ denotes the most significant bit of the cutoff $\mbk^*_j$ in the $j^{\text{th}}$ direction, $C\approx1.356$ is a constant, and $\Delta$ denotes the maximum derivative of the function over its whole domain when considered on a continuous interval, \ie an infinitely-resolved digitization. Crucially, assuming that the underlying distribution is sufficiently smooth, the bound becomes independent of $n_q$. Fig.~\SubFigRef{fig:reconstruction_fig}{c}) shows the number of bands required to reach a constant TVD of $\delta=0.04$ and $\delta=0.1$ as a function of system size $n$, for a range of gauge couplings $g=\{ 0.6,1,2 \}$. As the size is increased, the number of bands increases linearly (corresponding to an exponential increase in coefficients), until the critical threshold is hit, at which point the curves plateau, demonstrating an independence from system size.

\section{Measurement of Non-Stabilizerness on a Quantum Device}\label{sec:quantum_results}

Using the circuits approximating time evolution due to the Pauli-truncated Hamiltonian approach outlined in Sec.~\ref{sec:magic_numerics}, a second-order Trotter step at varying time steps is simulated on IBM's Heron r2 device, $\textbf{ibm\_marrakesh}$. The non-parallelizable CNOT depth is 51, which to the authors' knowledge is the deepest circuit where the non-stabilizerness has been measured on a quantum device. The number of Clifford circuits sampled ranged from $N_U=100$ to $N_U=200$. Running many circuits in parallel on a single chip was crucial for allowing economic use of the quantum resources. This required a carefully chosen layout, prioritizing high performing qubits. To choose the layout, the calibration parameters of the chip were queried before each run and the qubits were filtered based on having readout error less than $0.05$ and having connections to other qubits with two-qubit gate error rate less than $0.01$. Once the filtered set of qubits were chosen, contiguous patches of five qubits were found such that every patch was separated by at least one inactive qubit on the device. This approach minimized cross-talk between resonators.

The error mitigation methods used were Pauli Twirling (PT), measurement twirling, dynamical decoupling, and BSDR. For each individual circuit, the measurement twirls were averaged together, and then BSDR was applied to individual Pauli twirls, allowing for effective renormalization of $\ps$. Each shot of the physics circuit was Pauli twirled 28 times and then measurement twirled 5 times, where the number of measurements $N_M$ for each Pauli twirl were split evenly among the measurement twirls. The random Cliffords were omitted on the mitigation circuits so that a single set of mitigation circuit runs could be applied to all the physics circuits, allowing most of the quantum resources used to be applied to the physics circuits.

\begin{figure*}
    \centering
    \includegraphics[scale=.62, trim = 0 200 0 150]{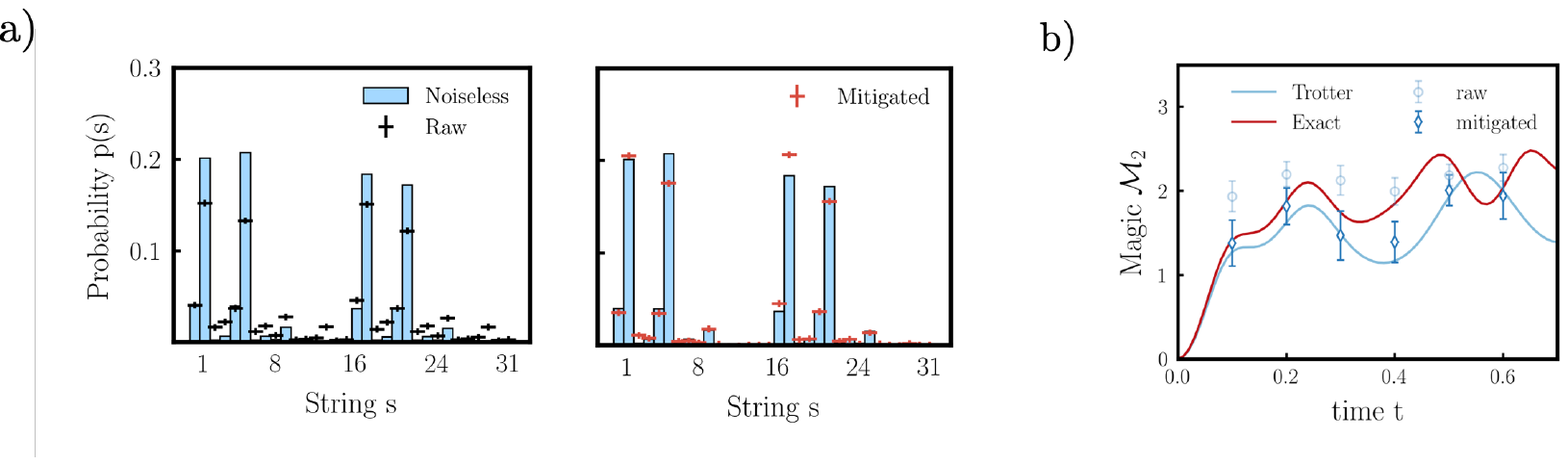}
    \caption{\textit{Quantum Measurement of non-Stabilizerness} \textbf{a)} Bitstring distributions for specific instance of random Cliffords. The raw data (left) shows small populations of many bitstrings whose populations are spuriously amplified by the noise of the quantum device. The bitstring distribution after application of BSDR (right) removes these spurious populations and restores much of the probability mass to the true bitstring. \textbf{b)} Values of non-stabilizerness over time. The solid lines correspond to evolution due to the exact propagator and the Trotterized circuit. The circles represent the calculations from the raw data and the diamonds represent the mitigated data. The raw data is almost constant, and the features of the dynamics are highly suppressed. The mitigated values are able to accurately track the real time dynamics for all times studied.}
    \label{fig:quantum_magic_fig}
\end{figure*}

Figure ~\SubFigRef{fig:quantum_magic_fig}{a}) (top) shows an example bitstring distribution from one particular random Clifford with the raw results and the mitigated results (where all the twirls are averaged over). The primary effect of the noise is to transfer probability from the most significant strings and distribute it among the less likely strings. This redistribution of probability can have a severe impact on the predicted value of non-stabilizerness from the device, and so applying the mitigation technique developed in this work is crucial. By mitigating the $\langle Z_k\rangle$ and inferring the distribution $\ps$ from the mitigated operators, the effects of the small but many erroneous bitstrings are suppressed and the most likely ones are amplified, restoring their significance.

The predictions of the non-stabilizerness are shown in Fig.~\SubFigRef{fig:quantum_magic_fig}{b}), and the numerical values are presented in Table~\ref{tab:reported_magic_values}. It is immediately apparent that without mitigation, the experimentally measured non-stabilizerness is higher than that predicted by classical simulation of the Trotter evolution, and the variation over time is suppressed. It is counterintuitive that the unmitigated non-stabilizerness is higher than that of the noiseless circuit in light of results on the complexity of noisy quantum circuits~\cite{schuster2025polynomial}. This effect can be understood by considering the Pauli spectrum of a state, \ie the set of all expectation values of Pauli strings. For stabilizer states on $n$ qubits, there will be $2^n$ Paulis that have expectation value plus or minus one, and the rest will be exactly zero. On the other hand, states with high non-stabilizerness generally have many nonzero Pauli expectation values with small magnitude. As previously observed, Pauli noise will reduce the magnitude of large expectation values and with finite measurement statistics, expectation values that should vanish exactly will be estimated to have a small but nonzero value. These effects taken together will cause the otherwise highly structured Pauli spectrum of a low magic state to resemble the more generic spectrum of a high magic state.
\begin{table}[h]
    \centering
    \begin{tabular}{ccccc}
    \toprule
    Time $t$ & Cliffords $N_U$ & raw & mitigated & Trotter \\
    \midrule
    0.1 & 100 & 1.893~(33) & 1.13~(23) & 1.271\\
    0.2 & 150 & 1.921~(37) & 1.58~(15) & 1.623\\
    0.3 & 150 & 1.923~(37) & 1.57~(15) & 1.487\\
    0.4 & 200 & 1.736~(28) & 1.29~(18) & 1.168\\
    0.5 & 100 & 2.030~(38) & 2.06~(14) & 1.93\\
    0.6 & 100 & 2.056~(42) & 1.85~(21) & 2\\
    \bottomrule
    \end{tabular}
    \caption{Raw, mitigated and Trotter values of $\mathcal{M}_2$ as a function of time. The raw values are determined by using the bitstring distributions that are read directly off of the device without any post-processing.}
    \label{tab:reported_magic_values}
\end{table}

\section{Conclusion}\label{sec:discussion}
Quantum simulation promises to answer questions about strongly-interacting many-body systems beyond the reach of current approaches. Non-abelian gauge theories are an appealing target, as their symmetries hinder analytic progress and large Hilbert spaces create computational barriers. This work improves the simulation of these theories in two ways: by investigating the quantum complexity of an SU(2) LGT in the mixed basis and by developing methods for efficient measurement of quantum information quantities on a quantum device.
First, the Stabilizer Rényi Entropy was calculated for a system of two interacting plaquettes. The non-stabilizerness of low-lying eigenstates was determined and its convergence with local Hilbert space dimension studied. Extrapolating from increasingly resolved digitization levels, $n_q=6$ qubits per plaquette were sufficient to reach one percent of the SRE value for low-energy eigenstates, which is 2-3 times the number of qubits needed for comparable precision in local observables like the energy~\cite{Froland:2025bqf}. It was also shown that different combinations of degenerate eigenstates can alter the required resources for simulating low-energy physics.
Second, randomized measurement schemes were improved by rewriting conventional estimators as a weighted sum over Clifford-averaged expectation values, reducing classical post-processing from $\mathcal{O}(16^n)$ to $\mathcal{O}(n2^n)$ via the Fast Walsh-Hadamard Transform. This work is the first application of this method to non-stabilizerness measurement protocols.

This connection between the non-stabilizerness estimator and operator expectation values motivated the central contribution of this work: a novel error mitigation scheme for bitstring distributions, called BSDR. The key insight is that common noise channels on current devices can be expressed as Pauli noise, which is more effectively corrected at the operator level. For states admitting a hierarchy in their Walsh spectrum, such as the low-energy subspace of field theories like the scalar field and SU(2) in the mixed basis, as well as states arising in applications such as PDEs and machine learning, only a constant number of observables is needed to reach a fixed error level in the reconstructed distribution.

BSDR was applied to real data from randomized measurements on IBM quantum devices, where raw results failed to capture time variations of non-stabilizerness. The mitigated results more accurately show the time dependency and the buildup of non-stabilizerness after a quench from a stabilizer state. This timescale is connected to quantum chaos and ergodicity~\cite{turkeshi2025magic,tirrito2024anticoncentration}, making BSDR suitable for studying non-equilibrium quantum dynamics on real devices through the lens of bitstring distributions~\cite{mark2024maximum}.

Extending this work to numerical studies of more plaquettes is an important future step. It has been recognized that the non-stabilizerness, in particular a basis-independent measure like the non-local magic, can serve as signals for the internal structure of bound particles. Understanding the implications of this for gauge fields is a promising direction~\cite{xu2026observation,cao2026string,jha2026magicentanglement11dimensionalsu2}. Furthermore, systematic studies comparing the non-stabilizerness in other bases like the electric representation and connecting this to the practical gate complexity and non-Clifford resources required in simulation are essential for reaching larger lattice volumes.

Another vital direction is determining how the Walsh spectrum of a state is related to other measures of complexity like non-stabilizerness and non-Gaussianity. It has been shown that for certain quantum states, a hierarchy in the non-Gaussian correlation structure allows for efficient representations~\cite{froland2025entanglement}; extending this concept to the realm of sequency truncation may allow for experimentally-friendly probes of otherwise complex observables. Additionally, understanding the underlying structure in the Walsh spectrum between states that are considered complex by different classes of resource theories may elucidate new phenomena in the context of quantum chaos and non-equilibrium dynamics.

Finally, it is expected that the BSDR error mitigation strategy presented in this work will find great use in many other simulation tasks relevant for high energy physics. As an example, many of the primitives that are used to construct wavepackets in scattering experiments, namely plane waves and Gaussian profiles, constitute smooth functions and so BSDR should prove to be beneficial~\cite{Zemlevskiy:2024vxt,Farrell:2025nkx,Lee:2026xfm}. Using BSDR will be advantageous for studying complex observables like the non-Stabilizerness, R\'enyi entropy, Inverse Participation Ratio, and other quantities that are strongly sensitive to device noise.

\begin{acknowledgments}
We gratefully acknowledge discussions with Jiunn-Wei Chen, Roland Farrell, Zhiyao Li, Martin Savage, Xiaojun Yao, and Nikita Zemlevskiy. Additionally, we thank Chris Kane and Nikita Zemlevskiy for useful comments on the draft.
This work was supported, in part, 
by U.S. Department of Energy, Office of Science, Office of Nuclear Physics, InQubator for Quantum Simulation (IQuS)\footnote{\url{https://iqus.uw.edu}} under Award Number DOE (NP) Award DE-SC0020970 via the program on Quantum Horizons: QIS Research and Innovation for Nuclear Science\footnote{\url{https://science.osti.gov/np/Research/Quantum-Information-Science}}.
It was also supported, in part, by the Department of Physics\footnote{\url{https://phys.washington.edu}}
and the College of Arts and Sciences\footnote{\url{https://www.artsci.washington.edu}} at the University of Washington. 
We acknowledge the use of IBM Quantum services for this work. The views expressed are those of the authors, and do not reflect the official policy or position of IBM or the IBM Quantum team. We have made extensive use of Wolfram {\tt Mathematica}~\cite{Mathematica},
{\tt python}~\cite{python3,Hunter:2007}, {\tt jupyter} notebooks~\cite{PER-GRA:2007} 
in the {\tt Conda} environment~\cite{anaconda},
and IBM's quantum programming environment {\tt qiskit}~\cite{qiskit}. 
\end{acknowledgments}

\emph{Note Added:} As this manuscript was in the final stages of completion, the authors were made aware of the recent work Ref.~\cite{jha2026magicentanglement11dimensionalsu2}. That work numerically studies, using classical computing methods, the non-stabilizerness in a 1+1 SU(2) LGT in the electric basis with a $j=1/2$ truncation with matter. It provides a complementary perspective to the numerical non-stabilizerness (quantum) studies in this work, focusing on large lattice volumes and low gauge field truncation, whereas this work focused on small lattices with a high gauge field truncation. Investigating the interplay of these two representations is an exciting future direction.

\bibliography{arxiv_magic}

\newpage

\appendix
\onecolumngrid  % This is the REVTeX-specific command
\clearpage

\section{The Non-Stabilizerness of Indicator States}\label{app:magic_indicator}
This section develops a toy model for understanding the behavior of non-stabilizerness in the ground state of the two-plaquette system. For a system of $dn$ qubits, the basis states are defined by a tuple of $d$ variables $\vec{x}=(x_1,\dots,x_d)$, each taking values in $\{1,\dots,2^n-1\}$. A cutoff $C$ is defined as $C=\sum_{j=0}^{n-1}C_j2^j$ where the coefficients $C_j$ denote the binary expansion of $C$. Given such a binary expansion of a cutoff, it is useful to define the set of bitstrings $S_j$ as those that lie within two adjacent powers of two, $j$ and $l$, \ie $S_j=\{x|2^j\leq x \leq 2^j+2^l\}$. The $d$-dimensional indicator state $\ket{\Theta_d(C)}$ is defined as
\begin{equation}
    \ket{\Theta_d(C)} = \sum_{\vec{x}}f(\vec{x})\ket{\vec{x}}, \qquad f(\vec{x})\propto
    \begin{cases}
        A_j\quad & \text{max}(\vec{x})\in S_j\;\;\\
        0 \quad &\text{else}
    \end{cases}
\end{equation}
where the proportionality denotes up to normalization. This state encodes an indicator function on basis states such that it takes value $A_j$ if the largest entry of $\vec{x}$ lie within two adjacent powers of two in the cutoff's binary expansion and vanishes otherwise. Solving for the non-stabilizerness of this general state can be complicated, but its structure can be constrained in a way that allows it to be exactly solvable.

First consider the one-dimensional state such that the cutoff has expansion $C=2^a+2^b$ with $a\geq b$, \ie there are only two powers of two in the cutoff's binary expansion. For $x_1\in S_1$ then the amplitude of the state is $A$ and if it lies within $S_2$ the amplitude is $B$, with $|S_1|=2^a$ and $|S_2|=2^b$. The normalization of this state is $\braket{\Theta_1}{\Theta_1}=A^22^a+B^22^b$. The non-stabilizerness of this state is found by considering the expression for $\mathcal{M}_2$ in terms of the amplitudes~\cite{tarabunga2024magic}
\begin{align}\label{eq:sre_estimator}
    e^{-\mathcal{M}_2} &= \sum_{\mbs_1,\mbs_2,\mbs_3,\mbs_4} c_{\mbs_1}c_{\mbs_2}c_{\mbs_3} c_{\tilde{\mbs}_4}c^*_{\tilde{\mbs}_1} c^*_{\tilde{\mbs}_2} c^*_{\tilde{\mbs}_3} c_{\mbs_4}^*
\end{align}
where
\begin{equation}
    \tilde{\mbs}_1=\mbs_2\oplus\mbs_3\oplus\mbs_4\quad
    \tilde{\mbs}_2=\mbs_1\oplus\mbs_3\oplus\mbs_4\quad
    \tilde{\mbs}_3=\mbs_1\oplus\mbs_2\oplus\mbs_4\quad
    \tilde{\mbs}_4=\mbs_1\oplus\mbs_2\oplus\mbs_3 \, .
\end{equation}
As the amplitudes are purely real, evaluation of the right hand side can be reduced to assessing for a given quadruplet of bitstrings $(\mbs_1,\mbs_2,\mbs_3,\mbs_4)$ the $S_j$ membership of the corresponding $(\tilde{\mbs}_1,\tilde{\mbs}_2,\tilde{\mbs}_3,\tilde{\mbs}_4)$. This membership only depends on which $S_j$ the original quadruplet is in and so for $J\leq n-1$ possible $S_j$ there are $\binom{J+3}{4}$ distinct cases that must be evaluated. For cutoff choice $C=2^a+2^b$ this reduces to 5 distinct cases, namely how many of the $\mbs_i$ are in $S_1$ and how many are in $S_2$, where the number of valid quadruples for each case being denoted as $t_i$. The bitstring representation of any $\mbs_i\in S_1$ can be any combination of $0$ or $1$ until the $a^{th}$ bit, whereas any $\mbs_i\in S_2$ will have a $1$ at the $a^{th}$ bit and have any combination of $0$ and $1$ until the $b^{th}$ bit and $0$ in all other positions. The counts $t_i$ are enumerated as follows:
\begin{enumerate}
    \item \textbf{None in $S_2$:}
    For any $\mbs_i\in S_1$, XOR-ing it with another $\mbs_j\in S_1$ will result in another string in $S_1$. There are $t_0=2^{4a}$ combinations of such strings.
    \item \textbf{One in $S_2$:}
    Consider the case where $\mbs_1\in S_2$, $\mbs_2,\mbs_3,\mbs_4\in S_1$. As with the previous case $\tilde{\mbs}_1$ will trivially be in $S_1$. For the remaining $\tilde{\mbs}_j$, the $a^{th}$ bit is $1$ due to $\mbs_1$ and so $\tilde{\mbs}_j$ is necessarily an element of $S_2$. This implies the bits in positions $b,\dots,a-1$ must be $0$, implying that the pair in $S_1$ must have bit patterns that cancel one another. The bits in positions $0,\dots,b-1$ are unfixed, and so the total number of configurations is
    \begin{align}
        t_1 &= 4\times2^b\cdot\sum_{patterns}(2^b)^3\nonumber\\
        &= 4\cdot 2^b\cdot 2^{a-b}\cdot 2^{3b}\nonumber\\
        &= 4\times2^{a+3b}
    \end{align}
    where the 4 accounts for the different ways to place the element of $S_2$ and the sum is over patterns in bit positions $b,\dots,a-1$.
    \item \textbf{Two in $S_2$:}
    Consider the case where $\mbs_1,\mbs_2\in S_2$, $\mbs_3,\mbs_4\in S_1$. This implies that $\tilde{\mbs}_3,\tilde{\mbs}_4\in S_1$ for any choice of $\mbs_1,\mbs_2$. By the logic of the previous case, for $\tilde{\mbs}_1,\tilde{\mbs}_2\in S_2$ the constituent bitstrings $\mbs_3$ and $\mbs_4$ must match in positions $b,\dots,a-1$. There are $2^{2b}$ choices for pairs of $\mbs_1$ and $\mbs_2$, and so $t_2=6\times2^{a+3b}$, where the factor of $6$ comes from all the ways of choosing two elements of the quadruplet to be in $S_2$.
    \item \textbf{Three in $S_2$:}
    Consider the case where $\mbs_1,\mbs_2,\mbs_3\in S_2$, $\mbs_4\in S_1$. All of $\tilde{\mbs}_1,\tilde{\mbs}_2,\tilde{\mbs}_3$ will trivially be in $S_1$. Further, $\tilde{\mbs}_4$ will always be in $S_2$ as $\mbs_1,\mbs_2,\mbs_3$ will always have $0$ in bit positions $b,\dots,a-1$. Therefore $t_3=4\times2^{a+3b}$ given all of the ways to choose the element of $S_1$ from the quadruplet.
    \item \textbf{All in $S_2$:}
    When all the $\mbs_i$ are in $S_2$, the $\tilde{\mbs}_j$ will trivially be in $S_2$ and so $t_4=2^{4b}$
\end{enumerate}

Collecting all of the contributing terms from the cases above and assigning them their corresponding factors of $A$ and $B$ gives the final expression for the non-stabilizerness $\mathcal{M}_2$ of
\begin{equation}\label{eq:indicator_magic}
    \mathcal{M}_2 = -\log\left(\frac{A^82^{4a}+14A^4B^42^{a+3b}+B^82^{4b}}{(A^22^a+B^22^b)^4}\right)
\end{equation}
As a special case of the above equation, consider when $a=b$ and $A=B$, which corresponds to a state with cutoff $C$ a power of two. The argument of the logarithm becomes unity and so the state is a stabilizer state. For $A\neq B$, these points will be minima. This suggests that the oscillations observed in the ground state of the two-plaquette system are an instance of the more generic behavior where states supported on  a power of two number of basis states are generally closer to stabilizer states.

Having considered the case of one dimension, the two-dimensional case follows straightforwardly. In particular, the computational basis quadruplets $(s_1,s_2,s_3,s_4)$ decompose into $x$ and $y$ directions such that the constraints act independently. Therefore, there are $25$ total cases, corresponding to all products of the five cases discussed in the one dimensional case. The number of contributing terms for each of the $25$ is just a product of the terms in the constituent one-dimensional cases. For cases $(2,2),(3,3),(4,4),(2,4),(4,2)$ there are different summands, depending on whether the strings in $S_2$ for the $x$ and $y$ directions fall on the same amplitude $c_{(s_y,s_x)}$ or different amplitudes. Summing all of the 25 cases gives the numerator
\begin{align}
\sum_{\mbs_1,\mbs_2,\mbs_3,\mbs_4} c_{\mbs_1}c_{\mbs_2}c_{\mbs_3} c_{\tilde{\mbs}_4}c^*_{\tilde{\mbs}_1} c^*_{\tilde{\mbs}_2} c^*_{\tilde{\mbs}_3} c_{\mbs_4}^* &\propto 2^{8a}A^8 
+ \left(14\cdot 2^{2a+6b} + 28\cdot 2^{5a+3b}\right)A^4B^4 \nonumber\\
&\hspace{30pt}+ \left(120\cdot 2^{2a+6b} + 48\cdot 2^{5a+3b}\right)A^2B^6 \nonumber\\
&\hspace{60pt}+ \left(2^{8b} + 28\cdot 2^{a+7b} + 2^{4a+4b+1} + 14\cdot 2^{2a+6b}\right)B^8
\end{align}
where the proportionality is up to the normalization of the state $\sqrt{4^aA^2+(2^{a+b+1}+4^b)B^2}$. Specializing to the case with $A=B$, the non-stabilizerness of this state can then be expressed as
\begin{equation}
    \mathcal{M}_2=-\log\left(\frac{2^{8a} + 76\cdot 2^{5a+3b} + 148\cdot 2^{2a+6b} 
+ 28\cdot 2^{a+7b} + 2\cdot2^{4a+4b} + 2^{8b}
}{(2^a+2^b)^8}\right)
\end{equation}
When $a=b$ the non-stabilizerness vanishes, whereas for any $b<a$ the non-stabilizerness will be non-zero. Therefore when the support of the indicator function is a power of two the state is stabilizer, and it is maximized when it is in-between a power of two, causing the oscillations in the non-stabilizerness of the ground state observed in the main body.

\section{Estimators for Stabilizer Entropies}\label{app:estimator_results}
\subsection{Estimator Derivation}
This section gives a complete derivation for the estimators given in the main text. Begin by considering the more general case of estimating the observable
\begin{equation}
    O^{(m)} = \sum_{\{\mbs_1\dots,\mbs_m\}}A^{-w(\oplus_i\mbs_i)}\prod_i\text{p}(\mbs_i)
\end{equation}
where $w$ is the weight of a bitstring and $\oplus_is_i$ denotes the element-wise XOR of all the $s_i$. The strategy will be to rewrite the kernel $A^{-w(\oplus_is_i)}$ as the Walsh-Hadamard transform of a set of undetermined coefficients $\{a_k\}$. Explicitly the ansatz is
\begin{equation}\label{eq:kernel_ansatz}
    A^{-w(\oplus_i\mbs_i)}=\sum_{\mbk}a_\mbk(-1)^{\mbk\cdot\left(\oplus_i\mbs_i\right)}
\end{equation}
and so all that remains is to determine the values of the $a_k$. Plugging this in to the expression for the estimator $O^{(m)}$ gives
\begin{align}
    O^{(m)} &= \sum_{\{\mbs_1,\dots,\mbs_m\}}\sum_{\mbk}a_\mbk(-1)^{\mbk\cdot\left(\oplus_i\mbs_i\right)}\prod_i\text{p}(\mbs_i)\nonumber\\
    &= \sum_{\mbk}a_\mbk\prod_i\sum_{\mbs_i}(-1)^{\mbk\cdot \mbs_i}\text{p}(\mbs_i)\nonumber\\
    &= \sum_{\mbk}a_\mbk\hat{\text{p}}_j^m(\mbk)
\end{align}
where $\hat{\text{p}}(\mbk)$ is the Walsh-Hadamard transform of the measurement distribution and in the second line we have factorized the product among the four replicas. Now all that is left is to determine the expression for the $a_\mbk$. This is done by taking the 
Inverse Walsh-Hadamard Transform of Eq.~\eqref{eq:kernel_ansatz}
\begin{align}
    a_\mbk &= \frac{1}{d}\sum_{\mbs}A^{-w(\mbs)}(-1)^{\mbk\cdot \mbs}\nonumber\\
    &= \frac{1}{d}\prod_b\sum_{s_b}A^{-s_b}(-1)^{k_bs_b}\nonumber\\
    &= \frac{1}{d}\prod_b\left(1+\frac{(-1)^{\mbk_b}}{A}\right)\nonumber\\
    &= \frac{1}{d}\left(1+\frac{1}{A}\right)^{n-w(\mbk)}\left(1-\frac{1}{A}\right)^{w(\mbk)}
\end{align}
where in the second line the sum over $s$ has been factorized into a sum over the individual bits. The observables $O^{(2)},O^{(4)}$ correspond to the cases $m=2,4$ respectively and $A=-2$. This gives the final expressions in the main text Eq~\ref{eq:decoupled_magic_estimators}.

\section{Walsh-Paley Based Reconstruction of Wavefunction Probabilities}\label{app:reconstruction_bounds}
\subsection{One Dimension}
% \subsubsection{One Dimension}
This section first describes the general grounds by which the Total Variational Distance (TVD) between the exact bitstring distribution and the reconstructed distribution may be bounded by the sequency spectrum of the underlying distribution. Then the dominant coefficients in each sequency band of relevant classical functions are calculated to give tight bounds on the reconstruction quality. Finally, different noise models are considered to understand how different types of noise might affect the reconstruction quality of the bitstring distributions.

To bound the TVD between two distributions, it is useful to view it as the $L_1$ norm of the probability vector $\ps-\psk{k^*}$. The $L_2$ norm of the vector is related to the Mean Squared Error (MSE)
\begin{equation}
    \varepsilon(\text{p}_1,\text{p}_2) = \sqrt{\sum_\mbs \left(\text{p}_1(\mbs)-\text{p}_2(\mbs) \right)^2}
\end{equation}
By an application of the Cauchy-Schwarz inequality the following bound applies
\begin{equation}
    \delta(\text{p}_1,\text{p}_2)\leq \frac{\sqrt{d}}{2}\varepsilon(\text{p}_1,\text{p}_2)
\end{equation}
where $d=2^n$ is the length of the vector for $n$ qubits. The MSE between the exact and reconstructed distributions is
\begin{align}
\label{eq:recon_err}
\sum_\mbs \left(\text{p}(\mbs)-\psk{\mbk^*}\right)^2 &= \sum_\mbs \left(\frac{1}{d}\sum_{\mbk > \mbk^*} (-1)^{\mbk\cdot\mbs}c_\mbk\right)^2\nonumber\\
&= \frac{1}{d^2}\sum_{\mbs}
\sum_{\mbk,\mbk' > \mbk^*}(-1)^{\mbs\cdot(\mbk+\mbk')}
c_\mbk c_{\mbk'}\nonumber\\
&= \frac{1}{d}\sum_{\mbk > \mbk^*}c_\mbk^2
\end{align}
where the third line used the orthogonality of Walsh functions. Therefore to bound the MSE it is crucial to understand the asymptotic decay of the sequency spectra. For convenience $\mbk^*$ will start at the beginning of the band $\nu$, thereby providing a worst case estimate for all elements of the band. In practice, starting at an arbitrary index in the block gives a small improvement to the error which is sub-leading to the improvement by including only the first element.

Under the assumption that the bitstring distribution is encoded by a smooth function $f(x)$, it has been shown that the Paley-ordered coefficients of degree $k$ obey the following bound~\cite{Yuen:1975abc}
\begin{equation}
    |c_\mbk|\leq F \times 2^{-q(\mbk)}
\end{equation}
where the prefactor $F=\max\limits_{x}|f^{(r)}(x)|$ is the maximum of the $r^{\text{th}}$ derivative of $f$ over the domain of $x$, $r=\sum_i\mbk_i$ is the number of ones in $\mbk$, and $q(\mbk)=\sum_ii\mbk_i+r(\mbk)$. The function $q(\mbk)$ obeys the following shift property $q(2^m+s)=(m+1)+q(s)$ relating bitstrings between blocks. The decay of the reconstruction error Eq.~\eqref{eq:recon_err} can therefore be expressed as
\begin{align}
    \sum_{\mbk>\mbk^*}\frac{1}{4^{q(\mbk)}} &= \sum_{j=m}^{n-1}\frac{1}{4^{j+1}}\sum_{k=0}^{2^{j}-1}\frac{1}{4^{q(k)}}\\
    &= \sum_{j=m}^{n-1}\frac{1}{4^{j+1}}\prod_i^{j}\sum_{k_i\in\{0,1\}}\frac{1}{4^{k_i(i+1)}}\\
    &= \sum_{j=m}^{n-1}\frac{1}{4^{j+1}}\prod_i^{j}\left(1+\frac{1}{4^{i+1}}\right)\\
    &\leq \frac{C}{3}\left(\frac{1}{4^m}-\frac{1}{4^n}\right)
\end{align}
where $C=\prod_i^n(1+1/4^{i+1})<1.356$ is a constant for $n\rightarrow\infty$; the first line represents the sum over different bands. The MSE is bounded by
\begin{align}
    \varepsilon^2(\text{p}_1,\text{p}_2) &\leq \frac{C\tilde{F}^2}{3d}\left(\frac{1}{4^m}-\frac{1}{4^n}\right)\\
\end{align}
where $\tilde{F}=\max\limits_{k}F$. Assuming that $n$ is large, the TVD dependence of the cutoff is
\begin{equation}\label{eq:tvd_final}
    \delta(\text{p}(\mbs),\text{p}(\mbs;\mbk^*))\leq  \max\limits_{r,s}|\text{p}^{(r)}(s)| \sqrt{\frac{C}{12}}\frac{1}{2^{m}}
\end{equation}
where $m$ denotes the most significant bit of the cutoff $\mbk^*$. Most notably, the only dependence on system size comes from the dependence on the derivatives. It is clear that for bitstring distributions that vary on the scale of individual bits, this will incur a factor exponential in system size. However for sufficiently smooth functions, namely those that have bounded derivative as $n\rightarrow\infty$, to achieve a constant TVD it is sufficient to calculate only the coefficients up to a particular band.

\subsection{Multiple Dimensions}
The case of encoding a D-dimensional function $f:\mathbb{R}^D\rightarrow\mathbb{R}$ on the interval $[0,1)^{\otimes D}$is now considered. In $D$ dimensions, the Paley-ordered Walsh functions are given by $W_{k}(x)=\prod^D_{j=1}W_{k_j}(x_j)$ where $k$ denotes a list of Walsh indices. It is useful to introduce the following functions
\begin{equation}
    Y(k,x)=\prod_jy(k_j,x_j),\quad y(k_i,x_i) = \frac{1}{(r(k_i)-1)!}\int_0^td\tau W_{k_i}(\tau)(x_i-\tau)^{r(k_i)-1}
\end{equation}
where $t,\tau\in\mathbb{R}^D$ and $k$ is a vector of bitstrings. The Walsh functions $W_k$ and $y$ satisfy the following properties:
\begin{enumerate}
    \item $\int_0^1(\prod_i t_i^{\alpha_i}W_{k_i})dt=0\;\;\text{if }\;\;\alpha_j<r(k_j)$\\
    \item $y(k_i,t)=\int_0^t\int_0^{t_1}\dots\int_0^{t_{r-1}}W_{k_i}(t_r)dt_r\dots dt_1$
    \item $\int_0^1y(k_i,t)dt=2^{-q(k_i)}$
\end{enumerate}
which are analogous to the single dimensional conditions given in \cite{Yuen:1972abc}.

 Denoting the block index of the band vector of $k$ as $\nu$, there are $\prod_{j=1}^{D} 2^{\nu_j-1}$ terms in the same band as $k$, defining a hypercube with corresponding side-lengths in $k$ space. 
 
 The bound on the Walsh coefficient $c_{k}$ is found by first expanding $f$ up to order $\rho=\sum_jr(k_j)$ around the boundary point $(1,\dots,1)$
 \begin{equation}
     f(x) = \sum_{|\alpha|<\rho}\frac{\partial_{\alpha}f}{\alpha!}(x-1)^{\alpha}+R_\rho(x),\quad R_\rho(x)=\sum_{|\alpha|=\rho}\frac{\rho(x-1)^{\alpha}}{\alpha!}\int_0^1 ds (1-s)^{\rho}D^{\alpha}f(1+s(x-1))
 \end{equation}
 where $\alpha=(\alpha_1,\dots,\alpha_D)$ is a multi-index of differentiation orders in each direction, $|\alpha|=\alpha_1+\dots+\alpha_D$, and $D^{\alpha}=\prod_i\partial_{\alpha_i}$. The term $R_r(x)$ is simply the integral form of the remainder for a Taylor expansion, and $t$ parameterizes a line from the boundary point 1 to the point $x$. The Walsh coefficient $c_{k}$ is found by integrating the series term by term against the Walsh function, where Property 1 causes the first terms to vanish.
 \begin{equation}
     c_{k}=\sum_{\alpha}\frac{\rho}{\alpha!}\int_0^1d\tau W_{k}(\tau)(x-1)^{\alpha}\int_0^1ds(1-s)^rD^{\alpha}f(1+s(x-1))
 \end{equation}
 The integral over $s$ is bounded by $\Delta=\max\limits_{x}f^{(\alpha)}(x)$, which is the maximum value of the directional derivatives specified by $\alpha$ over the integration region. Again by Property $1$, the only term in the sum over $\alpha$ that does not vanish is the combination of $\alpha$'s that matches the ranks of the corresponding index vector $\nu$. Using properties 2 and 3, the Walsh coefficient is then given by
 \begin{align}
     |c_{k}| &\leq \Delta\rho\prod_j \frac{1}{r(k_j)!}\int_0^1d\tau_j W_{k_j}(\tau_j)(x_j-1)^{r(k_j)}\\
     &= \Delta\frac{\rho}{\prod_jr(k_j)}\int_0^1d\tau Y(k,\tau)\\
     &\leq D\Delta\prod_j2^{-q(\nu_j)}
 \end{align}
 which is analogous to the bound in the one dimensional case with an extra factor of $d$ in front. Considering the case where the cutoff in each direction is $k_j^*$, the error will be given by summing over all $k$ outside this hypercube defined by $\mathcal{C}=\bigcup_jA_j$ where $A_j=\{k|k_j>k_j^*\}$. Applying a union bound to this sum gives the following
 \begin{align}
     \sum_{k\in\mathcal{C}}\frac{1}{4^{q(k)}} &\leq \sum_{j=1}^D\sum_{k\in A_j}\frac{1}{4^{q(k)}}\nonumber\\
     &= \sum_j\sum_{k_j>k_j^*}4^{-q(k_j)}\prod_{i\neq j}\sum_{k_i}4^{-q(k_i)}\nonumber\\
     &= C^{D-1}\sum_j\sum_{k_j>k_j^*}4^{-q(k_j)}\\
     &= \frac{C^D}{3}\sum_j\left(\frac{1}{4^{m_j}}-\frac{1}{4^n}\right)
 \end{align}
 where the final line follows from using the single variable sum over $j$ and $m_j$ is the most significant bit of the cutoff $k_j^*$. Once again, assuming large $n$, the bound on the TVD becomes
 \begin{equation}
     \delta\leq \sqrt{\frac{DC^D}{6}\Delta}\sum_j2^{-m_j}
 \end{equation}
 This bound is not tight in terms of its dependence on the dimension $D$. Crucially, for constant finite $D$ and a sufficiently smooth function, the bound is independent of the number of qubits used in the digitization.

\end{document}